# Security Threats and Research Challenges of IoT-A Review

*AKM Bahalul Haque\* and Sonia Tasmin*

Department of Electrical and Computer Engineering, North South University, Dhaka, Bangladesh

## ABSTRACT

Internet of things (IoT) is the epitome of sustainable development. It has facilitated the development of smart systems, industrialization, and the state-of-the-art quality of life. IoT architecture is one of the essential baselines of understanding the widespread adoption. Security issues are very crucial for any technical infrastructure. Since IoT comprises heterogeneous devices, its security issues are diverse too. Various security attacks can be responsible for compromising confidentiality, integrity, and availability. In this paper, at first, the IoT architecture is described briefly. After that, the components of IoT are explained with perspective to various IoT based applications and services. Finally, various security issues, including recommended solutions, are elaborated described and the potential research challenges and future research directions.

## 1. Introduction

The Internet of Things (IoT) has gained popularity in recent times. It is an interconnected network of devices like sensors, actuators, electronics, and software. A network or correlation among those gadgets helps to collect and share data between them. Each and everything can be defined uniquely utilizing an embedded computing device but can communicate within the current Internet infrastructure. IoT makes it possible to monitor and sense using the network infrastructure [1] to create opportunities for more effective physical incorporation into computer-driven networks. Besides, to minimize human interference, increase performance, precision, and economic benefit [2],[3] the IoT devices play a vital role. Internet of Things facilitates smart human living, sustainability, and a greener lifestyle. Moreover, IoT devices used in the industrial environment increase efficient product management through proper monitoring and risk management [4],[5].

IoT comprises sensors and actuators. It is an example of a broader class of cyber-physical networks involving intelligent grids, smart buildings, VPP (Virtual Power Plants), smart transport, and smart cities. It has a significant impact on the medical sector also. Among the applications, a wide range of equipment such as cardiovascular implants, biochip transponders for farm animals, cameras for broadcasting wild animal live feed in coastal waters, vehicles with embedded captors, environmental DNA analysis, food, surveillance for pathogens [6], or on-site operations supports firefighters in search and rescue operations [7]. IoT has spread its domain in every sector of socio-economic sectors. Legal scholars propose "thing" as a combination of hardware, software, information.

Similar to every other technology IoT has several issues regarding security and privacy. Since the IoT network is a combination of devices, communication technologies, and various protocols, security issues regarding availability, data integrity, data confidentiality, and authentication exist [8]. These issues hamper operational inefficiency, robustness, and throughput. For a sustainable and robust IoT network, security and privacy issues need to be adequately addressed. The reasons mentioned above can be a very impactful motivation for a comprehensive study regarding leveraging various issues.

Being IoT an impactful technology of recent times, it needs to be studied vigorously. Several pieces of research are going on for improving IoT and removing the security threats. Moreover, IoT has a tremendous impact on the industry and recent smart city improvement. Considering all the factors, it is indispensable to study IoT and perform critical research analysis, including contemporary literature. The analysis can be used to outline a sophisticated piece of literature that can help those trying to initialize their career in IoT and existing researchers looking for research gaps and current research challenges.

The rest of the paper is organized as follows-

Section 2 comprises an architectural analysis of IoT that includes various IoT layers. Section 3 consists of various IoT components that make the IoT system. An extensive analysis of security and privacy issues, including their state of the art recommended solution, is outlined in Section 4 and 5. In addition to the recommended solutions, Section 5 also comprises a recent literature analysis about IoT privacy and security issues. Section 6 outlines the future research directions that can be helpful for researchers and scientists. Finally, the paper concludes with a conclusion in Section 7.

## 2. IoT Architectures

Software integrated hardware devices process raw data and turn it into a usable format. Furthermore, the data is transmitted, stored, recovered, and analyzed with advanced IoT-integrated computer devices. Only a dependable IoT architecture layer can ensure a steady, durable, and swift connection between information and communication technology. Researchers have proposed several different architectures for the IoT environment. However, the three-layer structure is the most popular type among researchers and publications [9].

### 2.1 The Three-Layer Architecture

One of the primary and significant IoT architectures is the three-layer architecture. It is one of the most functional,

convenient, and easy to use architectures. The three-layers of this architecture are,

1. Application layer
2. Network layer
3. Edge/Perception layer

### 2.1.1 Application Layer

This layer defines all applications; no absolute norm is given. Its crucial function is to provide the customer with a particular service depending on the type of application. It can be seen in a variety of areas of IoT, such as applications for smart communities and homes, healthcare [10],[11], smart grids [12],[13], and automated vehicles [14],[15]. This layer may also work as a connectivity protocol, middleware [16], and cloud storage to enable server support. Therefore, security issues will vary depending on the context and industry of the application. In this specific architecture, various components are specified and focus on the IoT environment. We will need special binary programs or a special API (Application Programming Interface) on server-sides and client-sides [17]. The security architectures in most applications rely on the security of the DTLS CoAP protocol.

### 2.1.2 Network Layer

Its feature is to handle the transmitting and retrieving of information, with internet connectivity of different devices, among other layers. Besides, the network layer allows access to the edge/perception layer across various protocols and standards such as GPS, IEEE 802.X, and Near Field Communications (NFC). Internet protocols, cloud back-end networks, and smart devices support this layer [18]. Besides, the network layer can be managed based on the implemented environment with distinct aspects. However, Key Encryption Management and Systems, Intelligence Intrusion Detection Systems, and BlockChain technology [19],[20] constitute the most common network-level security framework in IoT architectures.

### 2.1.3 Perception or Edge Layer

The characteristics of the perception layer would be sensing abilities. IoT devices may communicate with customers or their working domains (sensors, smart meters, or IoT gateway edge-level servers). Those also collect environmental information with the help of smart objects. This layer undergoes multiple attacks due to the physical visibility of the edge layer in the IoT architecture. Safe channeling, an endpoint anti-malware solution, a multi-factor authentication system, and applications based on machine learning for cloud-based exception detection are the essential security components used in this layer [21],[22].

A wide range of IoT systems deficiencies has led to IoT devices' transformative use with computational capabilities in different application areas. In various disciplines, these limitations can create critical mistakes and data loss. In recent years, the protection of IoT ecosystems has also been identified as one of the trending issues that attracted the research society's attention [23].

### 3. IoT Components

Understanding the building blocks of IoT allows you to gain a deeper perspective of IoT's actual purpose and usefulness. We address six key elements required to carry out the functionalities of the IoT in the following parts.

### 3.1 Identification

For the IoT to align and rename facilities with their request, recognition is key. Many recognition mechanisms, such as ubiquitous codes (uCode) and electronic product codes (EPC) [24], are obtainable for the IoT. Furthermore, distinguishing between IoT objects and their addresses is essential. Object ID points to the object's name, for example, "T1" as a given temperature sensor. The address of the object corresponds to the address of the communication network. Besides, IPv6 and IPv4 provide the addressing methods of IoT objects. 6LoWPAN [25],[26] offers a compression mechanism for IPv6 headers, making IPv6 suitable for wireless networks with low capacity. It is imperative to differentiate between object identity and address because identification approaches are not globally unique, so addressing objects helps recognize those individually. Network objects within the range may also use public IPs instead of private ones. Identification techniques can be used for each object in the scheme to provide a particular identification.

### 3.2 Sensing

IoT sensing is a collection of data from linked items inside the network and a return to the data store, archive, or cloud. The collected data is analyzed for crucial decision-making purposes. The sensors used by IoT systems are wearable, smart actuators. Companies such as SmartStuff, Revolve, and Wemo, for instance, have smart hubs and mobile applications for thousands of smart devices and equipment to be monitored and controlled inside buildings via smartphones [27],[28]. In most IoT products (e.g., BeagleBone Black, Raspberry PI, Arduino Yun, etc.) A single-board computer (SBC) is embedded with sensors and incorporated protection features and IP/TCP. These systems typically bind to the central management portal to deliver relevant data to customers.

### 3.3 Communication

Heterogeneous objects are linked by IoT connectivity technologies to provide unique smart services. Usually, in the case of missing and noisy connections, low-power IoT nodes will operate. LTE-Advanced, Bluetooth, Z-wave, WiFi, and IEEE 802.15.4. provide networking protocols used by IoT. Some basic networking systems, such as Ultra-Wide Bandwidth (UWB), Near Field Communication (NFC), and RFID, are still used.

### 3.3.1 Communication Technologies

RFID (tags and readers) is the first technology used to incorporate the M2M principle. The RFID tag is a fundamental chip or tag that provides object identification. Furthermore, an RFID card reader sends a query signal or message to the tag, and the tag that is sent to the database gives a mirrored signal. The database based on reflector signals (10 cm to 200 m) [29]. The items are connected to the processing center.

The RFID transponders may be active, passive, semi-active, or semi-passive.

A battery drives active tags; here, passive tags are not needed. The control of the board is used when required for semi-passive / active labels.

The NFC Protocol supports up to 424 kbps of data with a high-frequency band of 13.56 MHz. When contact between active readers and passive tags or two active readers is established, the width can be 10 cm [30].

The UWB communication mechanism has been developed for low-level, low-energy, and high-bandwidth communications, which have recently improved their sensor connectivity capacities [31]. WLAN / WiFi, which uses radio waves for data sharing in a range of 100 m, is another networking technology [32]. In specific ad hoc environments, WiFi helps intelligent devices to link and share data without a modem. In short distances, Bluetooth is a networking system that uses short wave radio to relay data between devices to minimize energy consumption [33]. The Bluetooth-SIG (Bluetooth Special Interest Group) recently developed Bluetooth 4.1 that supports low-energy Bluetooth and high speed and IP networks [34].

### 3.3.2 Communication Protocols

For low-performance wireless communications aimed at extensible and safe networking, the IEEE802.15.4 specification defines all media and physical communication access [35]. The standard wireless connection through a GSM / UMTS network technology is originally LTE (Long-Term Evolution), based on high-speed data transfer from mobile phones [36]. It will protect high-speed devices and have multi-channel and transmitting services. LTE-A is an improved LTE version [37], with up to 100 MHz bandwidth, up and downlink space multiplexing, expanded coverage, increased latency, and decreased latency.

### 3.4 Computation

"The brain" and IoT's computing capabilities are control devices ( *e.g.*, FPGAs, microprocessors, microcontrollers, SOCs) and application software. Several hardware platforms have been developed to run IoT based applications. Besides, many application platforms are being used to have IoT capabilities. Operating systems (OS) are critical among such systems because they operate over the entire activation period. There are plenty of RTOS (Real-Time Operating Systems) that are excellent targets for RTOS-based IoT systems growth. To begin with, the Contiki RTOS was used extensively in IoT situations. Researchers and developers were aided by a simulator called Cooja (by Contiki) to simulate IoT and WSNs (Wireless Sensor Networks) [38].

Lightweight OS, Riot OS [39],[40], LiteOS, and TinyOS are also available for IoT environments. Many Google auto industry leaders founded the Open Auto Alliance (OAA). Furthermore, to step up the deployment of the IoV (Internet of Vehicles) model [41], they are aiming to make modifications to the Android version. Per the operating system has different characteristics. Another important computational aspect of IoT is Cloud Systems. These devices have the potential to move their data to the cloud with intelligent objects. This large amount of data can be analyzed in real-time to benefit the end-user. The host of IoT assistance is equipped with various free and commercial cloud systems and structures [42].

### 3.5 Services

Among all the IoT based services, identification programs are the most rudimentary and essential providers. Object detection is indispensable for an algorithm that brings real-life objects into the virtual world. Collaborative systems run in the Information Aggregation Systems background and use the collected information to evaluate and respond accordingly. Ubiquitous networks are therefore intended at all times and everywhere to provide Collaborative Aware Services. Both IoT implementations aim essentially to achieve a standard with universal services. Recent applications provide collaborative-aware services, information aggregation, and identity. Intelligent healthcare and smart grids come into the data collection group. Moreover, collective consciousness is closer to industrial automation, smart buildings, and smart transportation systems (ITS).

### 4. IoT Security and Privacy Issues

The IoT model involves addressing security flaws on various levels, including multiple applications and devices, from microchips to massive high-level computers. As mentioned below, we categorize the security risks surrounding the IoT deployment architecture,

- Low-level
- Intermediate-level and
- High-level

### 4.1 Low-level Security Concerns

As detailed below, the first protection level is concerned with safety problems in the data connection and physical layers of hardware and communication.

### 4.1.1 Sybil Low-level Threats and Spoofing

Sybil attacks are triggered by fraudulent Sybil nodes using false documents. A Sybil node will utilize arbitrarily fabricated MAC values to mask network resources as a separate unit on the physical level. Connection to infrastructure can then be declined to valid nodes [43].

### 4.1.2 Jamming Attacks

The jamming threats on wireless networks were directed at the weakening of the network through propagation without a clear radio waves specification. Radio disruption has a significant effect on network activities, resulting in failure or erratic actions by transmitting and receiving data by legitimate nodes [44],[45].

### 4.1.3 Attack of Sleep Privation

The energy-restricted IoT devices are vulnerable to attacks that lead to the sensor nodes remaining awake and "sleep loss". It contributes to battery failure as several activities in the 6LoWPAN setting are set to be carried out [46].

### 4.1.4 Insecure Start-up

A stable framework for IoT setup and configuration in the physical layer assures all devices' correct operation without infringing on privacy or network service interruption. The communication between the edge layer and the network layer must be protected against unauthorized access [47],[48].

### 4.1.5 Physical Interface Unreliable

Various physical conditions are associated with significant risks to the proper operation of IoT systems. Weak physical protection, software access through testing/debugging tools, and physical interfaces may impact network nodes [49].

### 4.2 Intermediate-level Security Concerns

Security problems at the intermediate level are directly associated with the routing, session management, and connectivity of the transport layers and IoT network, as mentioned below.

### 4.2.1 Sybil Attack

Sybil nodes can be added to degrade network efficiency and even violate data privacy, comparable to Sybil attacks on low-level layers. Sybil nodes can spam, disseminate malware, or trigger phishing attachments by interacting with a network's false identity. The network management system should authenticate all types of devices and users before logging in. Any network protection backdoor or wide security loopholes will expose the network to many vulnerabilities. The network is not safe. For example, the excess cost of Datagram Transport Level Security (DTLS) has to be reduced because of limited resources. The cryptographic methods to protect data transfer in IoT must consider the usefulness and lack of other tools [50],[51].

Message authentication protocols are very crucial for a successful and secure data transfer. As mentioned earlier, devices need to be authenticated with valid credentials. During data transfer, the route discovery process takes various phases, including address adjustment and router finding. The use of adjacent discovery packaging could have severe repercussions and denial of service without sufficient authentication [52].

### 4.2.2 Assault of RPL

The Lossy and Low-power Networks (RPL) IPv6 Routing Protocol is vulnerable to multiple attacks caused by the infected nodes. The attack will lead to resource depletion and deterioration [53]. Since a receiver node needs a buffer space to reassemble incoming packets, an attacker can be abused to deliver incomplete packets. This attack leads to denial of service because the space filled by the unfinished packets the attacker sends out is discarded for other fragment packets [54].

### 4.2.3 Fragmentation Repeat or Replication Assaults

Technologies adhering to the IEEE 802.15.4 specification, represented by limited frame dimensions, enable the convergence of IPv6 systems. The restoration of the 6LoWPAN layer of packet fragment fields could deplete capital, overflow buffer, and reboot the computers. The duplicate fragments sent via malicious nodes impact the reassembly and hamper other valid packets [55].

### 4.2.4 Sinkhole and a Wormhole Attack

The sinkhole attacks respond to routing requests by the attacker node, which allows the packet to travel the attacker node and then conduct a malicious operation on the network. The network attacks can further impair 6LoWPAN functions by wormhole attacks that build a tunnel between two nodes, causing bundles returning at a node to enter other nodes automatically. These attacks have significant effects, including denial of service, breach of privacy, and eavesdropping [56],[57],[58].

### 4.2.5 End-to-end Transportation Safety

The purpose of the transport-level end-to-end encryption is to provide a safe framework for efficiently retrieving the information from the sender node. Comprehensive authentication mechanisms are necessary to ensure the secure transmission of the message in encrypted form while ensuring minimal overheads.

Session hijacking with forged messages on the transport layer will result in denial-of-service. To begin the session between two nodes, the target node will be imitated by an invading node. The nodes will also need re-transmission by modifying the sequence numbers.

Different attacks that may breach location and identity protection can be seen on IoT's delay-tolerant networking (DTL) or cloud-based system. Likewise, an IoT deployment-based malicious cloud service provider may access sensitive information transmitted to the appropriate destination [59],[60],[61].

### 4.3 High-level Security Concerns

High-level security problems concern mainly the IoT applications, as mentioned below.

### 4.3.1 Unsafe Interfaces

The interfaces used by the internet, device, and cloud to access IoT resources are vulnerable to multiple attacks that seriously impact data privacy [49].

### 4.3.2 Security of Middleware

IoT middleware designed to make IoT paradigm interactions between heterogeneous organizations must be sufficiently secure to provide services. To ensure safe connectivity, various interfaces and environments use middleware [62],[63].

### 4.3.3 CoAP Security Issues

The high-level layer of the application layer is susceptible to attacks as well. A web transfer protocol for restricted computers, the Constrained Application Protocol (CoAP) uses DTLS connectors with different protection modes to ensure complete stability. To protect correspondence, the CoAP messages adopt a particular RFC-7252 format. Similarly, authentication and key management (AKM) are needed for the multicast support in CoAP [64],[65],[66].

### 4.3.4 Uncertain Software/firmware

Different IoT vulnerabilities include those triggered by insecure firmware/software. The code must be checked carefully for languages like JSON, XML, SQLi, and XSS. Similarly, firmwares must be updated securely and safely.

## 5. Recommended Solutions

IoT security threats target multiple elements that occur at all levels, such as firmware, network resources, physical equipment, software, applications, and interfaces. Users interact with the components via interfaces in IoT systems. Furthermore, their security mechanisms may even be dismantled. The protection threats countermeasures fix this communication's vulnerabilities in various layers to ensure a certain safety level. These countermeasures are further complicated by numerous protocols enabling component deployment. This section offers a summary of the critical safety strategies.

### 5.1 Low-level Pprivacy and Security Solutions

Some of the categorized security solutions are depicted as follows-

### 5.1.1 Anti-jamming Mechanism

The jamming attacks refer to interference that leads to communication conflicts or overflows for networks of wireless sensors. Young et al. are suggesting an approach to detect jamming attacks. The detection of attacks is possible by measuring the connection speed used for the set of vibration signals. These numbers are then evaluated to an adjusted optimum range for detection accuracy. By computing a successful packet delivery ratio, Xu et al. proposed to prevent jamming attacks by doing accuracy tests on signal intensity and the nodes' locations, the proposed algorithms work. Noubir et al. are considering another anti-jamming method using error-correcting codes and cryptographic functions. The system operates by splitting packets into blocks and interlines the encrypted packet bits. Likewise, spatial retreat and streaming techniques are recommended for coping with jam-attacks. Channel surfing allows legal channel frequency shift to contact devices. The space retreats, by comparison, allow specific devices to adjust their position at a certain distance when traveling to the target spot [67],[68],[69].

### 5.1.2 Safe Physical Layer Communication

Pecorella *et al.* suggest a system designed to ensure safe physical layer communication for the initialization of IoT. For the transmitted and receiving nodes, a low transfer speed is set to provide a missing eavesdropper. Other approaches to the implementation of artificial noise in signals are also used [70],[71],[72].

### 5.1.3 Detect Sybil Attack and Spoofing Threats

As a separate computer, a malicious Sybil node will use bogus MAC properties to masquerade as a different machine. That would lead to the loss of energy and the denial of connectivity to legitimate network equipment. Their strategy is used to evaluate the sender location by using tracker nodes during the message communication. Another message corresponds with the same sender location, but another user's identity is inferred as a Sybil attack. For detecting spoofing threats, other techniques by Li et al. and Chen et al. utilize signal intensity calculations for MAC addresses. Another Xiao et al. method involves channel prediction for detecting attacks from Sybil. The methodology uses multiple channel estimation identities and additional criteria to identify Sybil nodes [73],[74],[75].

### 5.1.4 Inappropriate Physical Protection

Devices with inappropriate physical protection are distinguished by external interfaces that offer access to firmware or applications and vulnerable utility tools for checking and debugging. Recommendations are issued by the Open Network Application Security Project (OWASP) to enhance IoT devices' physical security. Redundant hardware interfaces are essential to avoid. Debugging and testing methods must be removed. Hardware-based systems (e.g., Trusted Platform Modules) increases physical stability.

### 5.1.5 Sleep Deprivation Attacks

A system is developed to counteract wireless sensor sleep deprivation attacks. A cluster-based approach integrates the proposed structure, where each cluster is separated into many sectors. By eliminating long-distance communication, the consumption of electricity is minimized. With a five-layer architecture of the wireless sensors, the system performs intrusion detection. In the WSN model's upper layers, a cluster coordinator requires an expanded security mechanism and sink nodes and leader nodes. Similarly, in the lower levels of the WSN architecture, the follow-up nodes are fitted with basic intrusion detection systems [76].

### 5.2 Intermediate-level Privacy Solutions

Riaz et al. propose a safety system with device modules for secure data encryption, neighborhood discovery, authentication, and key generation. The elliptical curve encryption (ECC) is used for protected neighbor discovery. The ECC public key signatures are used in this process. Depending on the implementation specifications, both symmetric and asymmetric key management schemes are planned to be implemented. Data transfer across nodes happens in an encrypted manner to ensure confidentiality and integrity [77],[78].

For authenticating version numbers and ranks, the Authentication System called VeRA uses the Hash [79], MAC [80], and Digital Signature [81] Features. A rank and version number based authentication security service is proposed to mitigate adverse invasion while mapping through the IPv6 LLN (Low-Power and Lossy Network) routing protocol by Dvir et al. A lower parent node rank than the RPL norm requires the baby. No DAO messages are sent by the infected node, resulting in traffic delays by malicious nodes during transmission. A node's rank value can be reduced to find the root for eavesdropping [82],[83].

Zhou *et al.* [84] are aided in maintaining identification and privacy in a cloud-based IoT via a secure packet forwarding authentication method. The proposed architecture proposes an IoT network configuration from a central location for a hostile cloud service provider to protect an IoT network. Similarly, in the SMARTIE project, a forum for protecting data exchanged between IoT devices is suggested. Henze et al. are proposing a distributed platform for safe communication between IoT networks. Log message authentication is then used to denote hostile activity, which prevents cloud-based IoT from messages being changed, withheld, added, and reordered. To check across various gateways, it records control messages at several locations [85].

The RERUM project [86] suggests a system for Smart City IoT apps to ensure safety and stability. For IoT-based scenarios such as the smart healthcare sector and smart city, the project aims at validating trust and security. Similarly, for IoT environments, including smart communities, smart shopping, smart hospitals, and smart houses, the BUTLER project [87] advocates context-aware information systems. In the ARMOUR project [88], another mechanism for playing with security standards is introduced in an IoT base. The ARMOUR experiment determines defense design, creates testbeds, conducts tests, and produces qualification marks. As well as layer-specific safety specifications, the tests can be used to guarantee secure end-to-end communication. Lightweight cryptographic protocols were used in the project to enhance data security and integrity. Authentication-based techniques and Data integrity are being applied to build trustworthy applications.

### 5.3 High-level Privacy/security Solutions

Granjal *et al.* [89] proposed another solution to protecting messages for apps that connect via the web using different CoAP protection options. Brachmann *et al.* [90] suggest a solution that combines TLS and DTLS to stable CoAP-based Lossy and Low-power Network linked to the internet. Similarly, for IP networks, a security paradigm of 6LBR is proposed to filter messages and provide end-to-end security [91]. The SecurityEncap alternative uses the security options configuration and primarily performs the data transfer necessary for authentication and replay protection. TLS and DTLS routing is proposed to allow end-to-end protection that prevents LLNs from web-based threats.

A power-efficient security policy with a public-key authentication is suggested by Sethi *et al.* [92] for IoT-based CoAP. The proposed safety framework implemented by a test utilizes the Mirror Proxy (MP) and service directory that the server provides for sleep requests and a server (or endpoint) resource list. Project OWASP [93] lays out guidelines for IoT protection countermeasures. Protection protocols include configurations that check the interface against well-known bugs of the development tool (XSS and SQLi), use HTTPS and firewalls to deal with unsafe high-level interfaces, and discourage bad passwords.

Conzon et al. proposed the VIRTUS middleware that is used to protect distributed apps operating in an IoT system. The middleware uses a case-based connection method by integrating TLS and SASL for data integrity, XML stream encryption, and validation [94]. The authentication method guarantees resource protection and data sharing for registered users only. Integrated with network servers, the VIRTUS middleware helps in stable and flexible IoT applications being deployed. A semantic system called Otsopack [63] serves as a middleware to allow heterogeneous applications to communicate safely. Ferreira et al. are suggesting another protection architecture for IoT middleware. Liu *et al.* [63] propose a middleware server that promotes filtration of data during the connection between heterogeneous IoT systems. The standard features of authorization, authentication, and accounting are introduced via a critical hierarchy of app, root, and service keys. The proposed middleware enables an essential method for profiling, addressing, and naming through heterogeneous environments [95],[96].

### 5.4 Recent Critical Literature Contribution and Analysis

There has been a large number of studies during recent years. Some of the notable pieces of literature and their contributions are mentioned in a tabular format in Table 1 below-

Table 1 Recent Literature Contributions Related IoT Security and Privacy

| References | Year | Contribution |
|---|---|---|
| Dorri *et al.* [97] | 2017 | Assessing the requirements of IoT based smart city; Blockchain integration for security and privacy |
| Fremantle *et al.* [98] | 2017 | IoT and Blockchain integrated framework for IoT security threats |
| Oracevic *et al.* [99] | 2017 | Analysis of security issues and state of the art recommended solutions |
| Oh *et al.* [100] | 2017 | Comprehensive security analysis based on IoT elements; Proposed security requirements. |
| Ahemd *et al.* [101] | 2017 | Analysis of threats and countermeasures of various IoT layers; Assessing security providing technologies for addressing the risks. |
| Ouaddah *et al.* [102] | 2017 | Proposed blockchain and smart contract-based framework for IoT security |
| Salman *et al.* [103] | 2017 | Security and privacy issues analysis; Proposed a software model for securing IoT |
| Miraz *et al.* [104] | 2018 | Assessing blockchain-enabled cryptographic security mechanism for IoT Security; Depicting recent challenges faced while providing IoT security |
| Román-Castro *et al.* [105] | 2018 | Evaluating stateof the art security and privacy scenario and analyzing their prospects; |
| Vorakulpipat [106] | 2018 | In dept analysis and performance analysis of IoT architecture, applications, and various vulnerabilities and countermeasures |
| Roy *et al.* [107] | 2018 | In-depth analysis of blockchain and IoT architectures and their integration issues; Feasibility and possible integration analysis of blockchain for leveraging IoT security issues. |

| References | Year | Contribution |
|---|---|---|
| Xiao *et al.* [108] | 2018 | Feasibility assessment of implementing artificial intelligence against IoT security attack;<br>Various attack detection and secure authentication management using artificial intelligence. |
| Stergiou *et al.* [109] | 2018 | Integrating cloud computing and IoT ;<br>Proposed architecture for preventing security threats;<br>Efficiency and robustness analysis. |
| Sollins *et al.* [110] | 2019 | Big data related security and privacy attribute analysis;<br>Big data and IoT relationship assessment and propose design aspects addressing the security issues |
| Chaabouni *et al.* [111] | 2019 | Intrusion detection analysis of IoT networks for improving cyber defense;<br>Previous machine learning-based system development analysis during the recent past and addressing the future research challenges for IoT. |
| Nizzi *et al.* [112] | 2019 | Using HMAC for securing IoT and privacy protection;<br>In-depth analysis of the effect of address shuffling inside the entire network;<br>Proposed approach based on the result analysis. |
| Alraja *et al.* [113] | 2019 | Proposed framework for IoT based healthcare system usability;<br>User perception analysis towards the IoT based healthcare system usage, security, and privacy. |
| Hassija *et al.* [114] | 2019 | Comprehensive analysis of IoT based system application, security, and privacy analysis;<br>Various technology integration in IoT networks is assessed, including security and privacy issues. |
| Rahman *et al.* [115] | 2020 | Integrating blockchain in IoT;<br>Proposed SDN framework;<br>Addressing the security and privacy of IoT data; |
| Mohanta *et al.* [116] | 2020 | Analyzing blockchain for IoT security and privacy;<br>Analyzing IoT security threats;<br>Result-oriented case study analysis for the integration factors. |
| Dedeoglu *et al.* [117] | 2020 | Comprehensive, result-oriented and in-depth analysis of blockchain-based IoT security issues challenges and research directions;<br>Analyzing the opportunities and threats; |
| Hussain *et al.* [118] | 2020 | Analyzing the security attributes and threats of IoT;<br>Feasibility analysis of various artificial intelligence-based techniques and models for threat prevention; |
| Sharma *et al.* [119] | 2020 | Mobile IoT architectural analysis;<br>Security and privacy analysis in different layers and communication protocols;<br>Recent security privacy and implementation challenges are discussed briefly. |
| Mohanty *et al.* [120] | 2020 | Blockchain-based model for IoT privacy and security in the smart home environment;<br>Result analysis and performance comparison among the existing models. |
| Tewari *et al.* [121] | 2020 | Layered approach for threat and trust analysis in IoT;<br>Integration issues related to various IoT devices. |
| Islam *et al.* [122] | 2020 | Threat analysis of IoT based home systems;<br>Financial issues related to the home environment is discussed;<br>Blockchain-based approach in leveraging the problems; |

The above-mentioned table shows some of the recent literature related to IoT security and privacy. Apart from the mentioned points, the analysis can be depicted as follows-

- The IoT is studied extensively in recent times. Privacy and security issues are also discussed and analyzed in recent studies.
- IoT can be integrated with various other technologies. Researchers have made their approach to cloud computing and smart home-based techniques.
- Blockchain is one of the most promising technologies, and it is integrated with the internet of things. Blockchain can provide various facilities, for example, immutability, confidentiality, authenticity, and availability.
- IoT security and privacy issues can be addressed with blockchain technology. Though blockchain technology has several problems, such as scalability, interoperability, compliance issues, etc., the technology can be a potential white night for leveraging the security and privacy issues.
- Most recent studies related to IoT security and privacy involves blockchain. Scientists have been trying hard to find out various frameworks for addressing IoT security and privacy issues.
- Blockchain and IoT can be beneficial in potential research directions. If the application issues can be in-depth effectively, blockchain and IoT can be tools for developing smart and secure systems.
- There is a significant research gap in Mobile IoT device-related surveys and literature analysis. Extensive literature survey analysis can help analyze the implementation challenges, security, and privacy issue analysis, potentially finding potential research directions.

## 6. Challenges and Research Directions for the Future

From a privacy standpoint, blockchain application in the Internet of Things platforms and frameworks faces several obstacles. Researchers are incorporating blockchain into different IoT systems. This section addresses a few problems, open problems, and potential research paths from the perspective of confidentiality during the convergence of blockchain technology with numerous IoT implementations.

### 6.1 IoT in Industry

Due to its open and transparent existence, blockchain technologies in industrial IoT systems are growing. In a decentralized environment, for instance, in a production facility, IIoT detectors would be more efficient [123]. This is because, by updating the shared ledger at every stage, data can be spread to every single IIoT blockchain node. Many experiments have been carried out in previous literature to solve such privacy problems in IIoT systems, such as confidentiality and differential privacy, to maintain data integrity during industrial automation. However, before inclusion in the blockchain case, these methods need significant modifications. Therefore, such systems' privacy security is essential, and researchers should concentrate on protecting blockchain-based IIoT systems' privacy [124],[125].

### 6.2 Internet of Things for Farming

IoT based supply chain uses real-time monitoring of the production, manufacturing, shipment, housing, and distribution of agricultural goods. This traceability scheme aims to enhance farming and agrarian sector protection, supervision, cultivation, and processing practices. The monitoring and tracking processes in agriculture and agricultural IoT systems become more successful by using blockchain technologies. One such example is the leakage of any agricultural product's precise location and operation. Due to its diverse nature, intelligent contract security in blockchain-based IoT agriculture has enormous potential. Data leakage across the distribution cycle may be managed by writing successful codes based on secrecy. Future studies should propose combining privacy protection techniques in these systems by concentrating specifically on smart contracts and mixing strategies [126],[127].

### 6.3 Smart Cities

To further advance smart cities' ideas, researchers have begun combining blockchain with emerging smart city technology. Researchers have proposed that blockchain will remove multiple safety risks to smart cities due to its decentralized setting. Although blockchain is quite beneficial for smart communities, it often poses many privacy risks due to decentralization. Any hacker may enter the shared blockchain of a smart city and may attempt to acquire and infer sensitive details about smart city residents' personal lives and actions, resulting in significant privacy issues. Privacy security cannot only be grouped into a few predetermined domains in blockchain-based smart cities. However, for multiple smart city implementations, methods such as anonymization, smart contracts, and differential privacy can be used when the key prerequisite is to secure data sharing between different processes. Differential privacy is one of the possible choices, according to lightweight privacy protection in smart cities. It provides a reasonable guarantee of privacy, along with power over the utility of data as well [128],[129].

### 6.4 Crowd Sensing with Mobile Devices

A new sensing method called mobile crowdsensing has been introduced with the growing number of smart devices, exploiting smart device users' capacity, and gaining the advantage of using IoT technology for large-scale sensing. This transparency raises security issues for MCS apps. The crowd detector must provide clarity to MCS users while still transmitting data to the network in real-time. Blockchain-based crowdsensing systems need to guarantee that sensing by any effective privacy security process is anonymous and that no actual MCS user identities are exposed to adversaries. Using anonymization is one technique to protect the anonymity of MCS consumers. In this way, even though an adversary gets access to private data, the initial identities are not exposed. Noise in the data of MCS consumers using a differential privacy security approach may be another possible use. In a decentralized MCS environment, however, preserving the trade-off between precision and privacy can be difficult when users report their data in a real-time environment.

## 7. Conclusion

The article summarizes the interpretation of IoT architecture layers, the cooperation of IoT elements, and the applications of IoT. The internet has changed our way of living, shifting relations among people digitally in a couple of settings from intelligent life to social connections. IoT will probably apply another measure to this loop by empowering correspondence with and between smart items, subsequently prompting the vision of "whenever, wherever, whatever" communications using any media. Security issues are growing with the growth of IoT devices in many business areas and human lives. Because of the restriction in assets, a broad scope of weaknesses has developed. The more significant part of these weaknesses can prompt framework disappointment in the workplace of the IoT.

Furthermore, this paper critically analyses recent pieces of literature related to IoT security and privacy issues. The recommended solution analyzed in this paper provides state of the art overview of current cybersecurity situations of IoT. Recent literature analysis also shows the research areas to work on in the future so that this technology can reach its epitome. There will be many technological challenges for a resource-constrained system such as IoT, as mentioned throughout the paper. Similarly, with the advent of new technological innovation, there needs to be some solutions that can address the challenges. Some of the recommendations are mentioned in the paper, and others are yet to be implemented in the future.